\documentclass[aps,prl,showpacs,amsmath,amssymb,amsfonts,twocolumn,
superscriptaddress,floatfix,footinbib]{revtex4}

\usepackage[dvips]{epsfig}

\usepackage{subfigure}
\usepackage{graphicx}
\usepackage[flushmargin]{footmisc}
\usepackage{color}
\usepackage{hyperref}

\begin{document}

\title{Motion of spin polariton bullets in semiconductor microcavities}
 \author{C. Adrados}
    \affiliation{Laboratoire Kastler Brossel, Universit\'{e} Pierre et Marie Curie,
    \'{E}cole Normale Sup\'{e}rieure et CNRS, UPMC Case 74, 4 place Jussieu,
    75252 Paris Cedex 05, France}
  \author{T. C. H. Liew}
    \affiliation{Institute of Theoretical Physics, Ecole Polytechnique F\'ed\'erale
de Lausanne, CH-1015, Lausanne, Switzerland}

  \author{A. Amo}
   \affiliation{Laboratoire Kastler Brossel, Universit\'{e} Pierre et Marie Curie,
    \'{E}cole Normale Sup\'{e}rieure et CNRS, UPMC Case 74, 4 place Jussieu,
    75252 Paris Cedex 05, France}
    \affiliation{CNRS-Laboratoire de Photonique et Nanostructures, Route de Nozay,
91460 Marcoussis, France}

  \author{M. D. Mart\'in}
    \affiliation{Departamento de F\'isica Materiales and Instituto de Ciencias de Materiales 'Nicol\'as Cabrera', Universidad Aut\'onoma de Madrid,
Cantoblanco, E-28049 Madrid, Spain}

  \author{D. Sanvitto}
    \affiliation{Departamento de F\'isica Materiales and Instituto de Ciencias de Materiales 'Nicol\'as Cabrera', Universidad Aut\'onoma de Madrid,
Cantoblanco, E-28049 Madrid, Spain}
    \affiliation{NNL, Istituto Nanoscienze - CNR, Via Arnesano, 73100 Lecce,
    Italy}

  \author{C. Ant\'on}
    \affiliation{Departamento de F\'isica Materiales and Instituto de Ciencias de Materiales 'Nicol\'as Cabrera', Universidad Aut\'onoma de Madrid,
Cantoblanco, E-28049 Madrid, Spain}

  \author{E. Giacobino}
    \affiliation{Laboratoire Kastler Brossel, Universit\'{e} Pierre et Marie Curie,
    \'{E}cole Normale Sup\'{e}rieure et CNRS, UPMC Case 74, 4 place Jussieu,
    75252 Paris Cedex 05, France}
  \author{A. Kavokin}
    \affiliation{Laboratoire Charles Coulomb UMR 5221 CNRS-UM2
F-34095 Montpellier Cedex 5, France}
    \affiliation{Physics and Astronomy
School, University of Southampton, Highfield, Southampton, SO171BJ,
UK}

  \author{A. Bramati}
    \affiliation{Laboratoire Kastler Brossel, Universit\'{e} Pierre et Marie Curie,
    \'{E}cole Normale Sup\'{e}rieure et CNRS, UPMC Case 74, 4 place Jussieu,
    75252 Paris Cedex 05, France}
  \author{L. Vi\~na}
   \affiliation{Departamento de F\'isica Materiales and Instituto de Ciencias de Materiales 'Nicol\'as Cabrera', Universidad Aut\'onoma de Madrid,
Cantoblanco, E-28049 Madrid, Spain}

\pacs{71.36.+c, 71.35.Gg, 78.67.De}

\date{\today}

\begin{abstract}
\noindent The dynamics of optical switching in semiconductor
microcavities in the strong coupling regime is studied using time-
and spatially-resolved spectroscopy. The switching is triggered by
polarised short pulses which create spin bullets of high polariton
density. The spin packets travel with speeds of the order of $10^6$
m/s due to the ballistic propagation and drift of exciton-polaritons
from high to low density areas. The speed is controlled by the angle
of incidence of the excitation beams, which changes the polariton
group velocity.
\end{abstract}

\maketitle

{\it Introduction.---} Compact, solid-state, semiconductor systems
are amongst the most promising candidates for
the construction of optical signal processing devices (see ~\cite{Amo10} and references therein). 
Often it is desirable to hybridize the abilities of electronic
excitations with those of light, for example in semiconductor
microcavities, where exciton-polaritons arise from strong coupling
between cavity photons and quantum well excitons~\cite{Kavokin07}.
The strong polariton-polariton interactions yield rich
nonlinearities capable of creating individual optical
switches~\cite{Martin02,Lagoudakis02,Leyder2007,Amo10} and memory
elements~\cite{Gippius07,Paraiso10} with low threshold power. Their
short lifetime underlies a high speed ps dynamics, capable of
ultra-fast switching~\cite{Freixanet00,Shelykh2008,Amo09}.

Unfortunately, a problem arises when one considers how to link
multiple polariton-based switches into an optical circuit. Many
theoretical proposals rely on using ballistically propagating
polaritons to carry
information~\cite{Johne2010,Shelykh2009,Shelykh2010}. Here the short
polariton lifetime seems to become a drawback since it introduces a
signal loss mechanism, implying that some kind of signal
amplification is necessary. An alternative to ballistic propagation
is the use of 'polariton neurons' in a microcavity excited in the
bistable regime~\cite{Liew08}, where signals are carried by the
propagation of a domain wall separating regions in which the system
is locally in the low or high density stable states. The propagation
is expected to be confined along channels and the signal propagation
distance limited by the size of a background optical pump, oriented
at normal incidence, rather than the polariton lifetime.

Another important feature of polaritons is their spin degree of
freedom, which allows their application in spinoptronic devices;
polaritons exhibit one of two spin states ($\sigma^{+/-}$) that
couple to external circularly polarized light and can be used to
encode information instead of relying on intensity. Furthermore, the
Coulomb interaction between polaritons depends on their spin state.
It is usually assumed that the interaction constant between
polaritons with parallel spins is larger than the interaction
constant between polaritons with antiparallel
spins~\cite{Vina96,Vladimirova,Ciuti98,K-Kavokin05}. This allows
spin dependent switching to
occur~\cite{Amo10,Martin02,Leyder2007,Paraiso10,Lagoudakis02}.

In this work we merge the concepts of ballistic signal propagation
and polariton neurons, in the route towards the realization of the
recently proposed spin-optronic
devices~\cite{Johne2010,Shelykh2009,Shelykh2010}. Excitation with a
continuous wave ({\it cw}) optical pump at an angle supports the
triggering of a propagating bullet, which travels across the pump
spot for distances much longer than allowed by the polariton
lifetime, thus overtaking a drawback for the implementation of
realistic devices. The bullet corresponds to a temporary switch from
the low intensity ({\it off}) state to the high intensity ({\it on})
state in a highly non-linear regime. The signal propagation speed is
approximately given by the polariton group velocity, which is
controlled by variation of the optical excitation angle. Our
temporally and spatially resolved measurements are fully supported
by a theoretical model based on the Gross-Pitaevskii (GP)
equation~\cite{Shelykh2006,Liew08}.

{\it Experiment.---} Our experiment consists in switching a
polariton fluid, quasi-resonantly created in a GaAs microcavity
(with Rabi-splitting $5.1$ meV) with a {\it cw} Ti:Sa laser pump,
from the {\it off} to the {\it on} state by the application of a
weak triggering probe pulse (2 ps-long pulses every 12 ns). The pump
(FWHM 100 $\mu m$) and probe (FWHM 20 $\mu m$) beams impinge on the
microcavity, at a point where the cavity-exciton detuning is
approximately zero, with the same angle, the latter positioned in
such a way that the propagating polaritons are able to cross the
pump beam area from one side to the other \cite{Amo10}.

The {\it cw} pump has a Gaussian profile in space and is blue
detuned by $0.5$ nm from the lower polariton branch resonance, which
is at $837$ nm at the chosen point of the sample. The central
wavelength of the pulsed probe (duration of 2 ps, spectral width of
1.2 nm) matches the wavelength of the pump. We will start by
considering a circularly polarized pump and a co-circularly
polarized probe. In this case the system effectively simplifies to a
scalar system. As the pump power is increased, a nonlinear threshold
can be seen in the transmitted intensity \cite{Amo10,Adrados2010}.
Above threshold the whole transmitted beam has a top-hat-like
profile and is no longer Gaussian. Under these conditions bright
polariton bullets cannot be excited, so we instead choose the pump
power to be below the nonlinear threshold.

As shown in Ref.~\cite{Amo10}, a small \textit{cw} probe of enough
power, inside the large pump would result in the switch to the {\it
on} state of the whole pump spot. The region excited by the probe
will be the first to switch to the \textit{on} state. The ballistic
propagation of the injected polaritons to the neighbouring region
set by their propagation direction would then result in the
switching \textit{on} of the rest of the pump spot, in a relay
mechanism. Thus, we expect the propagation to take place at the
group velocity set by the in-plane wavevector of the pump. We can
test the validity of this mechanism using a pulsed probe. The
arrival of the probe will create a polariton bullet in the
\textit{on} state, which will propagate across the pump spot. In our
experiments we follow the movement of these \textit{on} polariton
packets with a synchroscan streak camera, gaining access to the
propagation speed.

    \begin{figure}[t]
        \centering
        \includegraphics[width=1\columnwidth]{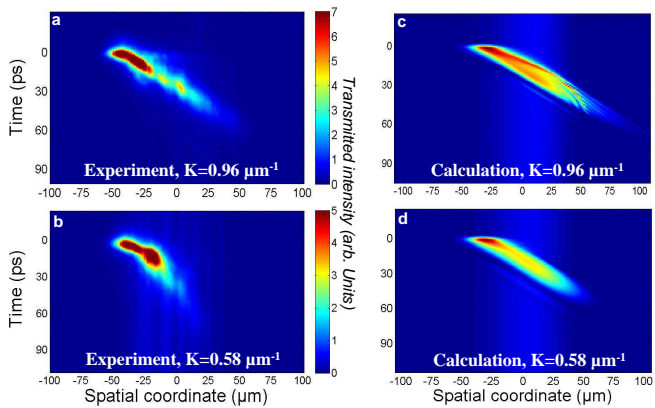}
    \caption{Streak camera (5 ps - time resolution) images of the near-field polariton intensity in space and
time for excitation wavevectors 0.96 $\mu m^{-1}$ (a) and
0.58 $\mu m^{-1}$ (b). Corresponding
theoretical images from the Gross-Pitaevskii equations are shown in (c) and (d).}
    \label{fig:2k}
    \end{figure}

Figures ~\ref{fig:2k}a and ~\ref{fig:2k}b display raw images from
the streak camera directly showing the motion of the polariton
bullet for two different in-plane wavectors: $0.96$ $\mu m^{-1}$ and
$0.58$ $\mu m^{-1}$ respectively. In the first case the bullet
propagates across the entire {\it cw} pump spot and for a time much
longer (the signal over noise ratio is bigger than 100 during 55 ps)
than the polariton lifetime ($\sim8$ ps in our sample). These
effects cannot be explained neglecting nonlinear interaction between
the pump and probe and considering the propagation of the probe
pulse only. For smaller in-plane wavevector, the bullet is more
difficult to distinguish since it does not propagate such a large
distance.

    \begin{figure}
        \includegraphics[width=1\columnwidth]{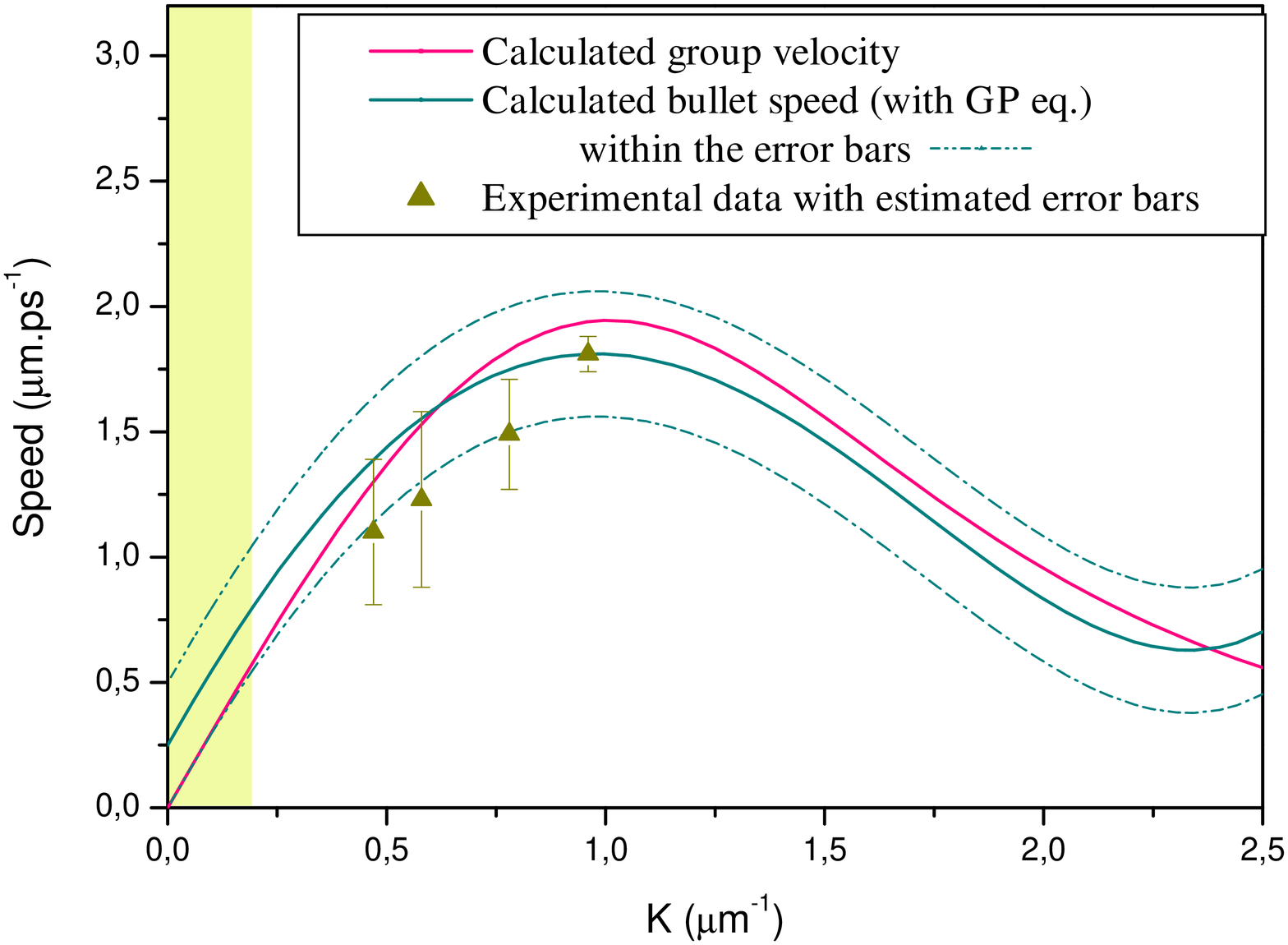}
    \caption{Dependence of the signal propagation speed on the in-plane wavevector.
Triangles show the experimentally measured points. The red curve shows the
polariton group velocity, $v_g(k)=\frac{1}{\hbar}\frac{d}{dk}E_\mathrm{LPB}(k)$. The solid blue curve
shows the signal propagation speed estimated from the solution of the
Gross-Pitaevskii equations. The dashed blue curves depict
the confidence bands of the estimation. The yellow part indicates a wavevector slot where the error bars are too big for the behaviour of the signal speed to be analyzed properly.}
    \label{fig:v_fct_k}
    \end{figure}

The slope of the emitted intensity shown in Fig.~\ref{fig:2k}a and b
can be used to estimate the signal propagation speed as
$1.81\pm0.07$ $\mu m.ps^{-1}$ and $1.23\pm0.35$ $\mu m.ps^{-1}$,
respectively. To first order, these values roughly match the
expected polariton group velocity as shown in
Fig.~\ref{fig:v_fct_k}.

    \begin{figure*}[t!]
        \includegraphics[width= 1\textwidth]{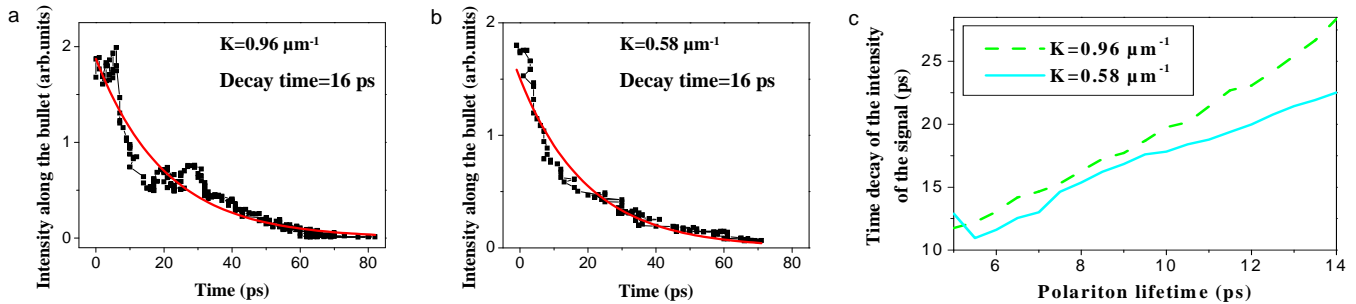}
    \caption{a)/b) Time dependence of the intensity (integrated over $\sim10^{7}$ single pulse realisations) of the polariton bullet
observed experimentally for $K=0.96$ $\mu m^{-1}$/$K=0.58$ $\mu
m^{-1}$. The solid red lines are exponential fits of the data. c)
The dashed/solid lines represent the theoretical dependence of the
polariton bullet decay time on the polariton lifetime for
K=0.96/0.58 $\mu m^{-1}$.}
    \label{fig:decrease_tail}
    \end{figure*}

Considering the size and shape of the pump beam at high power with
$K=0.96$ $\mu m^{-1}$ and the measured speed of the polariton
bullet, its lifetime matches the time to travel a $100$ $\mu m$
distance at $1.81$ $\mu m.ps^{-1}$. Note that the polariton lifetime
and pulse duration do not have a significant effect on the bullet
propagation speed. Nevertheless, they can play a role in the {\it
on} to {\it off} switching process. Indeed a given region remains in
the {\it on} state as long as the probe beam (directly or indirectly
via relay-polaritons) feeds it. Hence, when the polaritons of this
{\it on} region leave or decay via photon emission, the region falls
back to the {\it off} state provided that the pulse of the probe
beam is finished. Here we note that although the pump beam is
blue-detuned with respect to the lower polariton resonance and one
might expect to be in a bistable regime, due to the fact that
polaritons are propagating, there is no hysteresis. Essentially, the
state of the system in a given region in space no longer depends on
the history of that region, but rather on that of the neighboring
regions from which polaritons arrive.

The lifetime of the bullet can be measured by tracking its intensity
as it propagates. Figures ~\ref{fig:decrease_tail}a and
\ref{fig:decrease_tail}b depict the intensity decays corresponding
to Figs. \ref{fig:2k}a and b respectively. An exponential fit gives
a decay time, $\tau$, of about 16 ps, during which the bullets have
traveled $\sim30$ $\mu m$/$\sim20$ $\mu m$ for $K=0.96$ $\mu
m^{-1}$/$K=0.58$ $\mu m^{-1}$, after the trigger provided by the
probe pulse. Fig.~\ref{fig:decrease_tail}c, which displays the
calculated evolution (see model below) of $\tau$ with the polariton
lifetime, confirms that the lifetime of the bullets is almost
independent of the injected momentum and it is close to twice the
polariton lifetime for our experimental conditions.

An important feature in relation with the prospective use of this
system for high rate switching applications is the {\it on} to {\it
off} switching time at a given point in space. This time is as short
as 8 ps, and it is directly determined by the polariton lifetime.
Shorter switching times could be obtained in microcavities with
lower quality factors, resulting in shorter polariton lifetimes. 
The price to pay would be an increase of the switching power, as
higher powers would be required to generate and maintain the
polariton bullet propagation. Additionally, reducing the polariton
lifetime would lead to lower signal contrast. Therefore, a
compromise between a bigger switching rate and pump power is
essential.

{\it Theory.--- } The spatial dynamics of spinor coherent polaritons
can be described by the driven spinor GP
equations~\cite{Shelykh2006}:
\begin{align}
i\hbar\frac{\partial\phi_\sigma(\mathbf{x},t)}{\partial
t}&=-\left(\frac{\hbar^2\nabla^2}{2m_C}+\frac{i\hbar}{2\tau_C}\right)\phi_\sigma(\mathbf{x},t)+V\chi_\sigma(\mathbf{x},t)\notag\\&\hspace{20mm}+F_\sigma(\mathbf{x},t)\label{eq:GP_C}\\
i\hbar\frac{\partial\chi_\sigma(\mathbf{x},t)}{\partial
t}&=V\phi_\sigma(\mathbf{x},t)+\left(\Delta-\frac{i\hbar}{2\tau_X}+\alpha_1|\chi_\sigma(\mathbf{x},t)|^2\right.\notag\\&\hspace{20mm}+\alpha_2|\chi_{-\sigma}(\mathbf{x},t)|^2\left.\right)\chi_\sigma(\mathbf{x},t)\label{eq:GP_X}
\end{align}
where $\phi_\sigma(\mathbf{x},t)$ and $\chi_\sigma(\mathbf{x},t)$
represent the cavity photon and exciton wavefunctions, respectively,
which vary in the 2D plane of the microcavity depending on the
coordinate $\mathbf{x}$. $\sigma=\pm1$ is an index representing the
two possible circular polarizations of cavity photons, which couple
directly to excitons with spins represented by $\sigma=\pm1$,
respectively. $V$ represents the strength of this coupling. The
kinetic energy of cavity photons associated with their in-plane
motion is described by a parabolic dispersion with effective mass
$m_C$. We neglect the much higher effective mass of excitons and
denote the exciton-photon detuning at zero in-plane wavevector by
$\Delta$. Cavity photons are directly pumped by a coherent laser
described by the temporally and spatially dependent field
$F_\sigma(\mathbf{x},t)$, which is typically composed of a
superposition of \emph{cw} and pulsed pumps. Nonlinear interactions
take place between excitons, described by two constants, $\alpha_1$
and $\alpha_2$, which represent the strength of interactions between
excitons with parallel and antiparallel spins, respectively.

Starting from the initial condition that the wavefunctions are equal
to zero at all points in space, the dynamics of the system can be
calculated numerically from Eqs.~\ref{eq:GP_C} and \ref{eq:GP_X}.
This model has been proven to give a good description of
polarization sensitive nonlinear experiments in semiconductor
microcavities~\cite{Amo10,Leyder2007,Adrados2010}. Solving the GP
equations for excitation conditions corresponding to our experiment,
we obtain the signal propagation depicted in Figs.~\ref{fig:2k}c and
~\ref{fig:2k}d, which reasonably reproduce the experimental
behaviour. Using a similar procedure as that for the experimental
images, we can extract the dependence of the signal propagation
speed on the excitation wavevector, which is shown in
Fig.~\ref{fig:v_fct_k}. A clear correspondence can be found between
the signal propagation speed and the group velocity.

At low wavevectors it is hard to extract a signal propagation speed
since the distance travelled by the polariton pulse is minimal. We
note that the pump intensity and pump-lower polariton detuning was
kept fixed in the calculation of the curve shown in
Fig.~\ref{fig:v_fct_k}. The fixed pump intensity was chosen below
the nonlinear threshold intensity, in agreement with the
experimental excitation conditions. As the wavevector decreases the
nonlinear threshold increases due to changes in the photonic and
excitonic fractions, which decrease the effective polariton lifetime
and polariton-polariton interaction strength respectively. Close to
K=0 (yellow region in Fig.~\ref{fig:v_fct_k}) where the non-local
effects are less important, our calculations show a bistable
behaviour~\cite{Adrados2010} which is not accessible in our
experiments as they are performed and stored by averaging over
$\sim10^7$ single pulse realisations (the first pulse causes
permanent switching). In this case we should still observe the
motion of the switched {\it on} signal~\cite{Liew08}, but the
propagation speed is expected not to be given by the polariton group
velocity but by the gain in energy from the renormalisation
converted into kinetic energy. In this way, significant signal
speeds could be obtained even if the pump is oriented at normal
incidence.

{\it Spin-dependent switching.--- } Spin bullets of controlled
polarization can be prepared taking advantage of the spin dependence
of the polariton interactions. In this section we show that we can
follow the motion of a polariton bullet of a well-defined spin by
adjusting the polarization of the pump and probe beams.

    \begin{figure}
        \centering
        \includegraphics[width=\columnwidth]{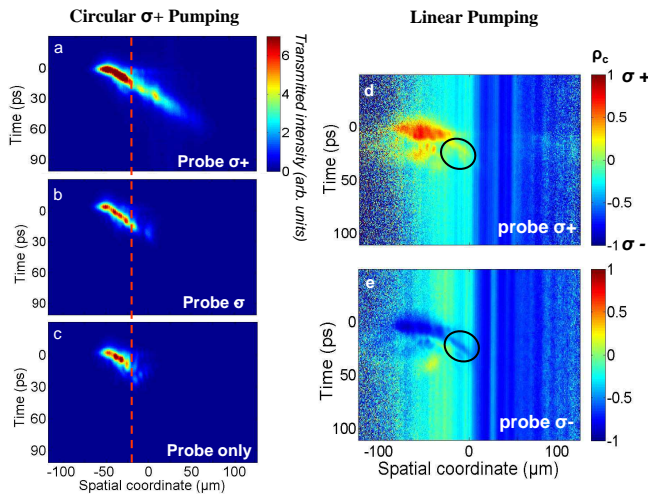}
    \caption{(a,b,c) Experimental propagation of a bullet when the pump is $\sigma^{+}$-polarized, with a $\sigma^{+}$ (a) or $\sigma^{-}$ (b) -polarized probe, and without the pump (c). The red line is a guide for the eye, evidencing the presence of the switch in the co-polarized case only. (d) and (e) show the degree of circular polarization $\rho_{c}$ obtained when the pump is almost linearly polarized ($\rho_{c}^{pump}=-0.2$), with a $\sigma^{+}$ (d) and $\sigma^{-}$ (e) -polarized probe. The black circles highlight the part of the signal arising from the switch, which is clearly $\sigma^{+}$ in (d) and $\sigma^{-}$ in (e).}
    \label{fig:pola}
    \end{figure}

Figures ~\ref{fig:pola}a and ~\ref{fig:pola}b show raw streak camera
images demonstrating the spin sensitivity in the switching of the
bullets. Comparing panel (a), where pump and probe polarizations are
both $\sigma^{+}$, with image (b), where they are cross-polarized,
we notice that in (a) the bullet propagates a much larger distance
and during a much longer time than in case (b). This demonstrates
that the switch only happens when the polarizations are the same.
This is further confirmed by comparison with panel (c) which shows
that the behaviour of the probe only is identical to that observed
in image (b) where polaritons behave as if the pump was absent and
spin up polaritons are not significantly affected by spin down
polaritons ($\left|\alpha_{1}\right|>>\left|\alpha_{2}\right|$).

In Figures ~\ref{fig:pola}d and ~\ref{fig:pola}e we have prepared
our \textit{cw} pump in a roughly linear polarization (ellipticity
given by $\rho_{c}^{pump}=-0.2$), and the pulsed probe in
$\sigma^{+}$ and $\sigma^{-}$ polarizations respectively. The false
colour scheme represents the degree of circular polarization
($\rho_{c}=(I^{+}-I^{-})/(I^{+}+I^{-})$, where $I^{+/-}$ is the
$\sigma^{+/-}$ right/left circularly polarized emitted intensity),
from -1 (blue, $\sigma^{-}$) to +1 (red, $\sigma^{+}$), while green
means $\rho_{c} = 0$, i.e. linear polarization. By adding a
$\sigma^{+}$ probe (d), we can observe that a $\sigma^{+}$ polarized
spin bullet propagates. A similar behaviour is observed by adding a
$\sigma^{-}$ probe to the linear pump (e), where now the spin bullet
is $\sigma^{-}$.

The reported behaviour under circularly polarised pump and probe
(conditions of Fig. 4a-c) can be understood in terms of a
polarization sensitive intensity switch, whose action propagates all
along the pumped area. This shows the great interest of this system
for the implementation of multiple gates integrated together in a
single semiconductor microcavity in order to make a device for
computation, analogous to those suggested in
Refs.\cite{Bouwmeester00,Wada04,Obrien09}.

{\it Conclusion.---} This work demonstrates the ultrafast motion of
spin bullets in semiconductor microcavities, opening the way for
building solid-state spin-optronic logic circuits, operating with
high speed and complete integrability. This is a significant step
towards realizing optical spin switches ~\cite{Amo10} and guided
transmission of information ~\cite{Liew08}.

We thank R. Houdr\'e for the microcavity sample. This work was
supported by the IFRAF, the \emph{Agence Nationale pour la
Recherche}, the Spanish \emph{MEC (MAT2008-01555,
QOIT-CSD2006-00019)}, \emph{CAM (S-2009/ESP-1503)} and \emph{FP7
ITN´s 'Clermont4' (235114)}, and \emph{Spin-optronics (237252)}.
A.B. is a member of the \textit{Institut Universitaire de France}.

\end{document}